\documentclass[prl,twocolumn,groupedaddress,nofootinbib]{revtex4}
\usepackage{latexsym}
\usepackage{amssymb}
\usepackage{amsmath}
\usepackage[hypertex]{hyperref}
\usepackage{graphicx}

\begin{document}


\title{ Grand Unification, Dark Matter, Baryon Asymmetry, and the Small
        Scale Structure of the Universe}

\author{Ryuichiro Kitano and Ian Low}
\affiliation{School of Natural Sciences, Institute for Advanced Study,
           Princeton, NJ 08540
}

\begin{abstract}

We consider a minimal grand unified model where the dark matter
arises from non-thermal decays of a messenger particle in the TeV
range. The messenger particle compensates for the baryon asymmetry in
the standard model and gives similar number densities to both the
baryon and the dark matter. The non-thermal dark matter, if
massive in the GeV range, could have a free-streaming scale in the
order of 0.1 Mpc and potentially resolve the discrepancies between
observations and the $\Lambda$CDM model on the small scale
structure of the Universe. Moreover, a GeV scale dark matter
naturally leads to the observed puzzling proximity of baryonic and
dark matter densities. Unification of gauge couplings is achieved
by choosing a ``Higgsino'' messenger.

\end{abstract}












\maketitle

\section{Introduction}

The standard model of
particle physics is believed to be incomplete. For
decades the strongest arguments are based more on aesthetic
reasonings than on empirical evidence. One example is
the fine-tuning in the mass of the scalar Higgs requires new
physics at around 1 TeV to stabilize the electroweak scale, for
which the benchmark solution is weak scale
supersymmetry (SUSY). Another example is the gauge
coupling unification which suggests a grand unified theory (GUT)
at high energy scale \cite{Georgi:1974sy}, assuming
a desert between the GUT scale and the electroweak scale where all
the particles come in complete multiplets of the GUT group. It
turned out that in the minimally supersymmetric standard model
(MSSM) \cite{Dimopoulos:1981zb}, gauge couplings
unify to a much better precision than in the standard model, which
adds to the attractiveness of the weak scale SUSY.

On the empirical side the situation has dramatically improved over
the last few years due to insights from
precision cosmological observations, including the existence of
dark matter, the acceleration of the cosmic expansion, the
baryon-asymmetric Universe, and a nearly scale-invariant density
fluctuations, none of which can be explained within the standard
model. (A minimal model addressing all these issues has been
proposed in \cite{Davoudiasl:2004be}.) Emerging from the 
observations is a description of the
Universe based on the $\Lambda$CDM model \cite{Spergel:2003cb}: a
tiny cosmological constant plus cold dark matter.

Neither the MSSM nor the $\Lambda$CDM model is perfect, however.
In the MSSM the most notable problems are the SUSY flavor problem,
new flavor violations from various superpartners, and the
non-observation of a light Higgs as well as any light sparticles,
which implies fine-tuning at a few percents level
\cite{Giusti:1998gz}. These problems prompted the proposal
\cite{Arkani-Hamed:2004fb} of giving up SUSY as a solution to the
hierarchy problem.
But if gauge coupling unification (and the neutralino dark matter)
is the main motivation for SUSY, there is a much simpler model,
the standard model with ``Higgsinos'', which is just as good
\cite{Arkani-Hamed:2005yv}.
On the other hand, there has been evidence suggesting, albeit not
yet conclusively, inconsistencies between observations and
numerical simulations of the $\Lambda$CDM model on the galactic
and sub-galactic scales \cite{Ostriker:2003qj}; it seems that
$\Lambda$CDM model predicts too much power on the small scales.
One way out is to introduce warm dark matter (WDM)
\cite{Colombi:1995ze,Bode:2000gq} which has a small free-streaming
scale $\lambda_{\rm FS} \lesssim 0.1\ {\rm Mpc}$, hence suppressing density
fluctuations on scales smaller than $\lambda_{\rm FS}$.

In this letter, using unification instead of naturalness as the
main incentive, we show that the three otherwise independent
aspects: gauge coupling unification, the puzzling proximity of
$\Omega_{\rm DM}$ and $\Omega_b$, and the small free-streaming
scale, can all be intertwined in a non-trivial way.
%
It is based on the model discussed in \cite{Kitano:2004sv} where
the dark matter arises from late-time decays of a heavy messenger
particle compensating for the baryon asymmetry in the standard
model.
%
%
%
First, gauge coupling unification suggests the existence of
``Higgsinos,'' which is the messenger particle, at the TeV scale
 and sets the GUT scale to be $\sim
10^{14}$ GeV \cite{Arkani-Hamed:2005yv}. Next cosmology constrains
the decay temperature of the ``Higgsinos'' to be $\sim 10$ MeV and
determines the scale of the higher dimensional operators
responsible for the Higgsino decay to be $10^{14-15}$ GeV
\cite{Kitano:2004sv}, a most natural value in the GUT picture.
Then the measured ratio of $\Omega_{\rm DM}/\Omega_b$ can be used
to fix the dark matter mass to be $\sim$ 1 GeV, which turns out to
come with the necessary free-streaming scale to suppress small
scale structures of the Universe. Alternatively, the need for a
sub Mpc free-streaming scale can be used to argue for a GeV scale
dark matter, with which the $\Omega_{\rm DM}$ is naturally close
to the $\Omega_b$.


\section{The Model}

The basic idea in \cite{Kitano:2004sv} is that the
dark matter $S$ is produced non-thermally from the decay of a
heavy messenger particle $X$, which carries the baryon number and
compensates for the baryon asymmetry in the Universe. Both $S$ and
$X$, which we call the dark sector, are assumed to be charged under 
a $Z_2$ symmetry, the $T$-parity, while the whole
standard model is $T$-even. The dark matter $S$ is then a stable
particle being the lightest $T$-odd particle (LTP). At the time of
baryogenesis, we assume that the $B-L$ number is distributed
between the $T$-even and $T$-odd sectors, resulting in the
following relation:
\begin{eqnarray}
 n_{B-L}^{\rm SM} = - n_{B-L}^{X} = - q_{B-L} ( n_{X} - n_{\bar{X}} ) \ ,
\label{bmnumber}
\end{eqnarray}
where $q_{B-L}$ is the $B-L$ charge of the messenger $X$, and
$n_{B-L}^{\rm SM}$ and $n_{B-L}^{X}$ are the $B-L$ number
densities in the standard model and the dark sector, respectively.
On the other hand, since both $X$ and $\bar{X}$ eventually decay
into the LTP, the dark matter candidate $S$, its number density is
given by the total number of $X$ and $\bar{X}$ particles
\begin{equation}
n_{\rm DM} = n_X^{\rm tot} \equiv n_X + n_{\bar{X}},
\end{equation}
which is independent of the $n_{B-L}^{\rm SM}$ in
Eq.~(\ref{bmnumber}) and would suggest there is {\it no}
connection between the baryonic and dark matter densities, unless
$n_X \gg n_{\bar{X}} \sim 0$ or the other way around. This implies
the lifetime of $X$ should be long enough so that it does not
decay until after most of the $\bar{X}$ particles annihilate with
$X$, which will be the case if there is no relevant or marginal
operator contributing to the decay of $X$. Then at temperature $T
< m_X$, where $m_X$ is the mass of the messenger, particle $X$
starts to annihilate with its anti-particle $\bar{X}$ through
gauge interactions and we are left with an abundance of $X$.
Consequently,
\begin{eqnarray}
 n_{B-L}^{\rm SM} = - n_{B-L}^{X} \simeq - q_{B-L}\  n_{\rm DM}\ .
\end{eqnarray}

This is a very general framework. In \cite{Kitano:2004sv} we worked out
the cosmological constraints, as well as an example
and the associated collider phenomenology. The model we are
interested here, is that with a singlet scalar dark matter $S$ and a
``Higgsino'' messenger $X$.
Recently it is pointed out \cite{Arkani-Hamed:2005yv} that
standard model plus Higgsinos achieves gauge coupling unification
at around $10^{14}$ GeV with an accuracy similar to that of the
MSSM.
At two-loop level, for $m_X=1$ TeV this minimal model predicts
smaller values of $\alpha_s$ than measured by ${\cal O}(5\%)$,
whereas larger values are predicted in the MSSM.
%
A GUT scale in the order of $10^{14}$ GeV is too small for a
conventional GUT model due to the constraint from proton decay. One
possibility, as discussed in \cite{Arkani-Hamed:2005yv}, is to embed the
setup in extra-dimensional or deconstructed models to lower the string
scale down to $10^{14}$ GeV or so. This has the bonus of resolving the
triplet-doublet splitting problem in conventional GUT models.
%
%
A crucial departure from the model in \cite{Arkani-Hamed:2005yv}
arises in that there the dark matter is a mixture of the
neutral component of the Higgsinos and an extra singlet fermion, 
which is the usual WIMP
scenario, whereas in our model the dark matter comes from
non-thermal decays of Higgsinos and has a mass which will be
determined to be in the GeV range.

The scenario proceeds in three stages. In the first stage,
baryogenesis is achieved by the out-of-equilibrium, CP-violating
decay of a $T$-odd scalar particle $P$ into $\bar{\ell}_i + X$ and
$\ell_i + \bar{X}$, where $\ell_i$ is the lepton doublet in the
$i$th generation in the standard model. CP violation enters
through the phases in the Yukawa couplings of the $P$ with
$\ell_i$ and $X$ if there are more than one $P$'s. So at one loop
one obtains an asymmetry proportional to the imaginary parts of
the Yukawa couplings of the $P$'s. After the $P$'s drop out of the
thermal equilibrium, we are left with an asymmetry in lepton
number in the standard model. Note that this mechanism is
reminiscent of the leptogenesis \cite{Fukugita:1986hr}, in which
$P$ is a heavy neutrino and $X$ is the ordinary scalar Higgs.
%
The effective $B-L$ number of $X$ in Eq.~(\ref{bmnumber}) is
determined to be $q_{B-L} = -1$.

The second stage is the complete annihilation of the Higgsino with its
anti-particles.
This constrains the mass of the messenger particle $m_X$ so that
$n_X \gg n_{\bar{X}} \sim 0$.
The thermally averaged cross section is
estimated to be $g^4/(256 \pi m_X^2 )$ for $s$-wave annihilations
into $SU(2)$ gauge bosons
which gives an upper bound on $m_X$:
\begin{eqnarray}
 m_X \ll 150 {\rm ~TeV} \left( \frac{1 {\rm ~GeV}}{m_S}
 \right)  \, .
 \label{bound-on-mX}
\end{eqnarray}
However, in order to establish the connection between the baryon
and dark matter number densities, $n_{B-L}^{\rm SM} \sim -q_{B-L}
n_{\rm DM}$, $m_X$ is preferred to be much lower than the
upper bound \cite{Kitano:2004sv}.
Moreover, a heavy $X$ spoils the gauge coupling
unification. 
%
We found that, increasing $m_X$ from 1 TeV to 100 TeV, the accuracy 
is worsened by ${\cal O}(1\%)$.
On the other hand, the lower bounds on $m_X$ simply come from direct
searches of new fermions which is about 100 GeV. Therefore it is quite
natural for the messenger particle to have a mass in the ${\cal O}$(TeV)
range.



The third stage is the late time decay of the Higgsinos into the dark
matter.
The decay has to occur before the big bang nucleosynthesis
(BBN), and after the completion of the pair
annihilation, as previously explained,
which requires the Higgsino to survive long after it falls out of
thermal equilibrium:
\begin{eqnarray}
\label{tdbounds}
  10~{\rm MeV}  \lesssim \, T_d  \, \ll \frac{m_X}{20}  \ ,
\end{eqnarray}
where $T_d$ is the temperature at which $X$ decays.
%
%
%
%
The long lifetime can be realized by assuming that $X$ only decays through a
dimension five operator
\begin{equation}
{\cal O}_{\rm decay} = \frac1M (X S)(H e^c_{i}), \label{odecay}
\end{equation}
where $H$ is the scalar Higgs and $e^c_{i}$ is the right-handed
charged lepton in the $i$th generation.
The above bounds on $T_d$ translate into bounds on $M$:
\begin{equation}
  10^{10}~{\rm GeV}
  \left( \frac{m_X}{1~{\rm TeV}} \right)^{1/2}  \ll M
\lesssim 10^{14} {\rm ~GeV}  \left(\frac{m_X}{\rm 1~TeV}
\right)^{3/2} .
\label{eq:M-upper}
\end{equation}
If we assume that the particle responsible for baryogenesis, $P$,
is thermally produced, the reheating temperature of the Universe
is restricted to be
\begin{equation}
 m_X \le m_P \lesssim T_R \lesssim
1~{\rm TeV}
\left(
\frac{M}{10^{14}~{\rm GeV}}
\right)^2
\ ,
\label{trbounds}
\end{equation}
where the upper bound comes from the requirement that the
thermally produced $S$ through the operator in Eq.~(\ref{odecay})
is less significant than the non-thermal component.
Eq.~(\ref{trbounds}) gives a stronger lower bound on $M$: 
$10^{14}\, {\rm GeV}\, (m_X/1 {\rm TeV})^{1/2}$.
Therefore, to be consistent with Eq.~(\ref{eq:M-upper}), we need
$m_X \gtrsim 1$ TeV. For example, if $m_X \sim 10$ TeV, $M$ can be
$\sim 3 \times 10^{14-15}$ GeV and the reheating
temperature could be $10-10^3$ TeV.
We emphasize that the lower bound in Eq.~(\ref{trbounds}) depends
on the assumption that the $P$ particle is thermally produced. In
a scenario where $P$ dominates the energy density of the Universe
and then decays into $X$ and radiation, $T_R$ can be much smaller
than $m_X$. Then $m_X$ is bounded below only by direct searches.

From the discussions above, we see that it is quite natural to
identify $M$ with $M_{\rm GUT} \sim 10^{14}$ GeV.
In this case, the decay temperature $T_d$ is just before the BBN,
$T_d \sim 10$ MeV, for $m_X \sim 1$~TeV.
A priori it is not clear at all that the cosmological bounds on 
$M$ should be consistent with the GUT scale $\sim 10^{14}$ GeV in our model.
For example, if the messenger were to decay through a dimension six 
operator, then the bounds on $M$ would lie between $10^8$ and $10^9$
GeV \cite{Kitano:2004sv}, which would make the identification with $M_{\rm GUT}$ less 
convincing.


There are in fact two dangerous marginal operators allowed by
$T$-parity: ${\cal
O}_4 = \bar{\ell}_i X S$, which contributes to the Higgsino
decay, and ${\cal O}_4' = SS H^\dagger H$, which could thermally
produce $S$.
The first one, being a Yukawa-type coupling, will be absent if we
set the initial value at $M_{\rm GUT}$ to zero, which we will
do. The second one can be fine-tuned away,
since we are not motivated by naturalness principle. In fact, if
we extend the model a little bit by using a complex scalar for $S$,
then ${\cal O}_4$ is forbidden by a $Z_3$ symmetry under which both
$S$ and $X$ have unit charge. Then
the Higgsinos decay through $\bar{X}\ell S S$. (Another 
dangerous operator,
$\bar{X} \ell S^\dagger$, can be forbidden by a $Z_2$ symmetry under 
which $X$ is even and $S$ is odd.) In addition, 
${\cal O}_4'=S S^\dagger H H^\dagger$ is generated radiatively with 
a coefficient suppressed 
by lepton Yukawa couplings, thus alleviating the fine-tuning. 
In any case, the scalar
masses, the Higgs and the dark matter, are fine-tuned.

Note that even though $X$ decays into dark matter plus leptons,
the excess in lepton number doesn't get converted into the baryon
number because the decay always happens after the electroweak
phase transition \cite{Kitano:2004sv} and the sphaleron process is
ineffective. We then obtain the number densities $n_B$ and $n_{\rm
DM}$ as follows:
\begin{eqnarray}
 n_B =  \epsilon \ n_{B-L}^{\rm SM} \ ,\ \ \
 n_{\rm DM} = | n_{B-L}^{\rm SM}|\ ,
 \label{eq:relation}
\end{eqnarray}
where the efficiency $\epsilon$ is the relation between the $B-L$
number and the baryon asymmetry in the presence of the sphaleron
process. It is different from the standard model value
\cite{Harvey:1990qw}, now that the presence of $X$ modifies the
charge neutrality conditions, and calculated to be $25/79$ with
the additional constraint that the total $(B-L)$ number is zero
before the decay of $X$. Therefore the ratio of the baryonic and
dark matter densities $\Omega_{\rm DM}/\Omega_b$ is given by
\begin{eqnarray}
\label{central} \frac{\Omega_{\rm DM}}{\Omega_b}= 3.16\
\frac{m_{\rm DM}}{m_p}  ,
\end{eqnarray}
from Eq.~(\ref{eq:relation}), where $m_p$ is the proton mass.
The measured ratio is $\Omega_{\rm DM}/\Omega_b\sim 5.1$ which implies
$m_{\rm DM}\sim 1.6\, m_p = 1.5$ GeV.
Eq.~(\ref{central}) is a central prediction of our model: the
baryonic mass density in our Universe will be close to the dark
matter density if the dark matter has a mass close to the nucleon
mass.
Indeed, we will see that considerations from reducing the small scale
structure of the Universe prefers light dark matter, 
${\cal O}({\rm GeV})$, showing a non-trivial consistency in this 
framework.
%

\section{Cosmology}

The cosmology with a dark matter candidate arising from
non-thermal decays is very interesting. In the $\Lambda$CDM model
the dark matter decouples from the thermal bath while being
non-relativistic and has a velocity distribution peaked at
around zero
at the time of structure formation. In other words, the momentum
profile is Maxwellian. On the other hand, the non-thermal dark
matter (NTDM) is relativistic when produced through decays of the
messenger particle and its momentum distribution is peaked at
$m_X/2$, which subsequently gets
red-shifted by the expanding Universe and becomes non-relativistic.

Such a primordial velocity dispersion has important implications
in structure formation, as the dark matter can smooth out
inhomogeneities by streaming out of overdense regions and into
underdense regions, the so-called Landau damping
\cite{Kolb:1990vq}, which is characterized by the free-streaming
scale $\lambda_{\rm FS}$. By definition cold dark matter has a
small $\lambda_{\rm FS}$ that is irrelevant for structure
formation, whereas hot dark matter, $\lambda_{\rm FS} \gtrsim 40$ Mpc,
has such a large free-streaming scale that would prevent galaxy
formation in the early epoch. So far observations on large scale 
structures
 clearly prefers a dark matter candidate that is cold, and
hence the $\Lambda$CDM model. Nevertheless, as mentioned,
on small scales (sub Mpc) simulations of CDM seems to
be at odds
with observations. First, the substructure
of CDM halos is predicted to be richer
than observed. Second, the simulated density profiles of dark matter
halos are generally cuspier than inferred from rotation curves. At this
moment it is not clear if this is indeed a failure of the $\Lambda$CDM
model, as there are complex issues with large N-body and hydrodynamics
simulations.

These discrepancies do not diminish the tremendous successes of the
$\Lambda$CDM model on larger scales, however, if these problems are
real, they present a great opportunity. The simplest known mechanism
for smoothing out small scale structures is the Landau damping.
In this regard a
popular candidate is the WMD, hot dark matter cooled down, which has
a free-streaming scale \cite{Bode:2000gq}
\begin{equation}
\lambda_{\rm FS} = 0.2\ (\Omega_{\rm WDM} h^2)^{\frac13} \left(
        \frac{m_{\rm WDM}}{\rm keV} \right)^{-\frac43}\ {\rm Mpc} .
\end{equation}
Observed properties of Lyman $\alpha$ forest constrain the power
spectrum at small scales and therefore put a lower bound on the
mass of the WMD \cite{Narayanan:2000tp}: $m_{\rm WDM} \ge 750$
eV, which translates into $\lambda_{\rm FS} \le 0.16$ Mpc for
$\Omega_{\rm WDM} h^2 = 0.15$. (A more recent analysis puts a
weaker bound: $m_{\rm WDM} \ge 550$ eV \cite{Viel:2005qj}.)
Furthermore, high-resolution cosmological N-body simulations 
seem to find better
agreements with observations for $\lambda_{\rm FS} \sim 0.1$ 
Mpc \cite{Avila-Reese:2000hg}.
For comparison, we can calculate
the free-streaming scale for the NTDM in our case \cite{Kolb:1990vq},
\begin{eqnarray}
\lambda_{\rm FS} &=& \int_{t_{\rm decay}}^{t_{\rm EQ}}
 \frac{v(t)}{a(t)} dt\ \simeq \  \int_{0}^{t_{\rm EQ}}
 \frac{v(t)}{a(t)} dt \\
 &\simeq& 0.16 \left(\frac{m_X}{3~\rm TeV}\right)
       \left(\frac{\rm GeV}{m_S}\right)
 \left(\frac{10~\rm MeV}{T_d} \right) \ {\rm Mpc}, \nonumber
\end{eqnarray}
where $v(t)$ and $a(t)$ are the physical velocity and the FRW scale
factor, respectively, of the NTDM at time $t$, and $t_{\rm EQ}$ denotes
the time for matter-radiation equality. 
Therefore, for a prototypical
$m_X/m_S/T_d =$ 3 TeV/1 GeV/10 MeV scenario that is well-motivated from
previous discussions, the NTDM is able to produce
enough power on the small scales to be consistent with the Lyman
$\alpha$ forest data and potentially resolve the discrepancies mentioned
above. There are also constraints
coming from studies of phase space density \cite{Hogan:2000bv}, which are
weaker than those coming from Lyman $\alpha$ forest \cite{Lin:2000qq}.

On the other hand, the NTDM is not exactly the same as the
WDM, since they still have different momentum distributions. As an
example, power spectrum of non-thermal production of neutralinos
by the decay of topological defects was considered in \cite{Lin:2000qq}
and found to be different from that of a 1 keV WDM 
for $k > 5\, h\, {\rm Mpc}^{-1}$. 

\section{Summary}

Using gauge coupling unification and cosmology as the main hints for 
physics beyond the standard model, we have considered a minimal GUT model 
which attributes a common origin to the baryon asymmetry and the 
dark matter, giving similar number densities to both the dark matter and 
the baryon. The dark matter, a singlet scalar, is produced by non-thermal
decays of a messenger particle, the Higgsinos, whose existence implies 
the unification at $\sim 10^{14}$ GeV. There are checks on the 
the model from several orthogonal directions. 
Cosmological bounds point to a Higgsino in the TeV range and a 
mass scale consistent with the GUT.
The ratio  
$\Omega_{\rm DM}/\Omega_b$ can be used to determine the mass of the dark 
matter to be GeV, which in turn gives a sub Mpc free-streaming scale
consistent with observations and, at the same time, has the potential of
resolving the $\Lambda$CDM crisis by reducing the power spectrum 
on small scales. Cosmology may be the best arena to test this model. The 
Higgsinos, on the other hand, might be too heavy to be discovered in the
coming collider experiments.


\vspace{0.2cm} \noindent {\bf Acknowledgments}
We benefited from conversations with Neal Dalal and Carlos Pe\~na-Garay.  
This work is supported by
funds from the IAS and in part by the DOE grant number DE-FG02-90ER40542.



\begin{thebibliography}{nn}

\bibitem{Georgi:1974sy}
H.~Georgi and S.~L.~Glashow,
Phys.\ Rev.\ Lett.\  {\bf 32}, 438 (1974);
H.~Georgi, H.~R.~Quinn and S.~Weinberg,
Phys.\ Rev.\ Lett.\  {\bf 33}, 451 (1974).

\bibitem{Dimopoulos:1981zb}
S.~Dimopoulos and H.~Georgi,
Nucl.\ Phys.\ B {\bf 193}, 150 (1981);
S.~Dimopoulos, S.~Raby and F.~Wilczek,
Phys.\ Rev.\ D {\bf 24}, 1681 (1981).

\bibitem{Davoudiasl:2004be}
  H.~Davoudiasl, R.~Kitano, T.~Li and H.~Murayama,
  arXiv:hep-ph/0405097.



\bibitem{Spergel:2003cb}
See, for example, D.~N.~Spergel {\it et al.}  
Astrophys.\ J.\ Suppl.\  {\bf 148}, 175 (2003)
[arXiv:astro-ph/0302209].


\bibitem{Giusti:1998gz}
L.~Giusti, A.~Romanino and A.~Strumia,
Nucl.\ Phys.\ B {\bf 550}, 3 (1999) [arXiv:hep-ph/9811386].

\bibitem{Arkani-Hamed:2004fb}
N.~Arkani-Hamed and S.~Dimopoulos,
arXiv:hep-th/0405159.

\bibitem{Arkani-Hamed:2005yv}
  N.~Arkani-Hamed, S.~Dimopoulos and S.~Kachru,
  arXiv:hep-th/0501082.


\bibitem{Ostriker:2003qj}
 For a review, see J.~P.~Ostriker and P.~J.~Steinhardt,
  Science {\bf 300} (2003) 1909
  [arXiv:astro-ph/0306402].


\bibitem{Colombi:1995ze}
S.~Colombi, S.~Dodelson and L.~M.~Widrow,
Astrophys.\ J.\  {\bf 458}, 1 (1996) [arXiv:astro-ph/9505029].

\bibitem{Bode:2000gq}
P.~Bode, J.~P.~Ostriker and N.~Turok,
Astrophys.\ J.\  {\bf 556}, 93 (2001) [arXiv:astro-ph/0010389].

\bibitem{Kitano:2004sv}
R.~Kitano and I.~Low,
Phys.\ Rev.\ D {\bf 71}, 023510 (2005) [arXiv:hep-ph/0411133].





\bibitem{Fukugita:1986hr}
  M.~Fukugita and T.~Yanagida,
  Phys.\ Lett.\ B {\bf 174}, 45 (1986).

\bibitem{Harvey:1990qw}
J.~A.~Harvey and M.~S.~Turner,
Phys.\ Rev.\ D {\bf 42}, 3344 (1990).


\bibitem{Kolb:1990vq}
  E.~W.~Kolb and M.~S.~Turner,
  ``The Early Universe,''
Redwood City, USA: Addison-Wesley (1990).


\bibitem{Narayanan:2000tp}
  V.~K.~Narayanan, D.~N.~Spergel, R.~Dave and C.~P.~Ma,
  arXiv:astro-ph/0005095.

\bibitem{Viel:2005qj}
  M.~Viel, J.~Lesgourgues, M.~G.~Haehnelt, S.~Matarrese and A.~Riotto,
  arXiv:astro-ph/0501562.

\bibitem{Avila-Reese:2000hg}
P.~Colin, V.~Avila-Reese and O.~Valenzuela,
  arXiv:astro-ph/0009317;
  V.~Avila-Reese, P.~Colin, O.~Valenzuela, E.~D'Onghia and C.~Firmani,
  arXiv:astro-ph/0010525.

\bibitem{Hogan:2000bv}
  C.~J.~Hogan and J.~J.~Dalcanton,
  Phys.\ Rev.\ D {\bf 62}, 063511 (2000)
  [arXiv:astro-ph/0002330].

\bibitem{Lin:2000qq}
  W.~B.~Lin, D.~H.~Huang, X.~Zhang and R.~H.~Brandenberger,
  Phys.\ Rev.\ Lett.\  {\bf 86}, 954 (2001)
  [arXiv:astro-ph/0009003].


\end{thebibliography}
\end{document}